\documentclass[11pt]{article}
\usepackage[utf8]{inputenc}
\usepackage[T1]{fontenc}
\usepackage{graphicx}
\usepackage{grffile}
\usepackage{longtable}
\usepackage{wrapfig}
\usepackage{rotating}
\usepackage[normalem]{ulem}
\usepackage{amsmath}
\usepackage{textcomp}
\usepackage{amssymb}
\usepackage{capt-of}
\usepackage{hyperref}
\usepackage{tikz}
\usepackage{xcolor}
\usepackage{authblk}
\usepackage[cm]{fullpage}
\usepackage{cite}
\usepackage{breqn}
\allowdisplaybreaks
\usepackage{mathtools,amsmath,amssymb,amsfonts,dsfont,mathrsfs,amsthm,latexsym}

\usepackage{bm}
\usepackage{graphicx}
\usepackage{centernot}
\usepackage{xcolor}
\usepackage{comment}
\usepackage{feynmf}
\usepackage{siunitx}
\usepackage{array}
\newcolumntype{L}[1]{>{\raggedright\let\newline\\\arraybackslash\hspace{0pt}}m{#1}}
\newcolumntype{C}[1]{>{\centering\let\newline\\\arraybackslash\hspace{0pt}}m{#1}}
\newcolumntype{R}[1]{>{\raggedleft\let\newline\\\arraybackslash\hspace{0pt}}m{#1}}
\usepackage{tikz}
\usepackage{pgfplots}
\pgfplotsset{width=7cm,compat=1.14}
\usepackage{braket}
\usepackage{xparse}
\usepackage{breqn}
\usetikzlibrary{shapes,
  positioning,
  arrows,
  decorations.pathmorphing,
  decorations.markings,
  calc,
  shadows.blur,
  shadings}
\usepackage[framemethod=tikz]{mdframed}

\makeatletter
\pgfdeclaredecoration{gluon}{coil}
{
  \state{coil}[switch if less than=%
    0.5\pgfdecorationsegmentlength+
    \pgfdecorationsegmentaspect\pgfdecorationsegmentamplitude+%
    \pgfdecorationsegmentaspect\pgfdecorationsegmentamplitude to last,
               width=+\pgfdecorationsegmentlength]
  {
    \pgfpathcurveto
    {\pgfpoint@oncoil{0    }{ 0.555}{1}}
    {\pgfpoint@oncoil{0.445}{ 1    }{2}}
    {\pgfpoint@oncoil{1    }{ 1    }{3}}
    \pgfpathcurveto
    {\pgfpoint@oncoil{1.555}{ 1    }{4}}
    {\pgfpoint@oncoil{2    }{ 0.555}{5}}
    {\pgfpoint@oncoil{2    }{ 0    }{6}}
    \pgfpathcurveto
    {\pgfpoint@oncoil{2    }{-0.555}{7}}
    {\pgfpoint@oncoil{1.555}{-1    }{8}}
    {\pgfpoint@oncoil{1    }{-1    }{9}}
    \pgfpathcurveto
    {\pgfpoint@oncoil{0.445}{-1    }{10}}
    {\pgfpoint@oncoil{0    }{-0.555}{11}}
    {\pgfpoint@oncoil{0    }{ 0    }{12}}
  }
  \state{last}[next state=final]
  {
    \pgfpathcurveto
    {\pgfpoint@oncoil{0    }{ 0.555}{1}}
    {\pgfpoint@oncoil{0.445}{ 1    }{2}}
    {\pgfpoint@oncoil{1    }{ 1    }{3}}
    \pgfpathcurveto
    {\pgfpoint@oncoil{1.555}{ 1    }{4}}
    {\pgfpoint@oncoil{2    }{ 0.555}{5}}
    {\pgfpoint@oncoil{2    }{ 0    }{6}}
  }
  \state{final}{}
}

\def\pgfpoint@oncoil#1#2#3{%
  \pgf@x=#1\pgfdecorationsegmentamplitude%
  \pgf@x=\pgfdecorationsegmentaspect\pgf@x%
  \pgf@y=#2\pgfdecorationsegmentamplitude%
  \pgf@xa=0.083333333333\pgfdecorationsegmentlength%
  \advance\pgf@x by#3\pgf@xa%
}
\makeatother

\tikzset{
  boson/.style={decorate,decoration={gluon,segment length=9pt,aspect=0}},
  on each segment/.style={
    decorate,
    decoration={
      show path construction,
      moveto code={},
      lineto code={
        \path [#1]
        (\tikzinputsegmentfirst) -- (\tikzinputsegmentlast);
      },
      curveto code={
        \path [#1] (\tikzinputsegmentfirst)
        .. controls
        (\tikzinputsegmentsupporta) and (\tikzinputsegmentsupportb)
        ..
        (\tikzinputsegmentlast);
      },
      closepath code={
        \path [#1]
        (\tikzinputsegmentfirst) -- (\tikzinputsegmentlast);
      },
    },
  },
  mid arrow/.style={postaction={decorate,decoration={
        markings,
        mark=at position .5 with {\arrow[#1]{stealth}}
      }}},
}
\tikzset{
    vector/.style={decorate, decoration={coil,aspect=0}, draw},
	provector/.style={decorate, decoration={coil,aspect=0,amplitude=2.5pt}, draw},
	antivector/.style={decorate, decoration={coil,aspect=0,amplitude=-2.5pt}, draw},
    fermion/.style={draw=black, postaction={decorate},
        decoration={markings,mark=at position .55 with {\arrow[draw=black]{>}}}},
    fermionbar/.style={draw=black, postaction={decorate},
        decoration={markings,mark=at position .55 with {\arrow[draw=black]{<}}}},
    fermionnoarrow/.style={draw=black},
    gluon/.style={decorate, draw=black,
        decoration={coil,amplitude=4pt, segment length=5pt}},
    scalar/.style={dashed,draw=black, postaction={decorate},
        decoration={markings,mark=at position .55 with {\arrow[draw=black]{>}}}},
    scalarbar/.style={dashed,draw=black, postaction={decorate},
        decoration={markings,mark=at position .55 with {\arrow[draw=black]{<}}}},
    scalarnoarrow/.style={dashed,draw=black},
    electron/.style={draw=black, postaction={decorate},
        decoration={markings,mark=at position .55 with {\arrow[draw=black]{>}}}},
	bigvector/.style={decorate, decoration={coil,aspect=0,amplitude=4pt}, draw},
}

\DeclareDocumentCommand{\Ag}{ s o }{ \IfBooleanTF{#1}
    { \IfValueTF{#2}{ \bm{\mathcal{A}}_{(#2)} }{ \bm{\mathcal{A}} } }
    { \IfValueTF{#2}{    {\mathcal{A}}_{(#2)} }{    {\mathcal{A}} } } }
\DeclareDocumentCommand{\Af}{ s o }{ \IfBooleanTF{#1}
    { \IfValueTF{#2}{ \boldsymbol{A}_{(#2)} }{ \boldsymbol{A} } }
    { \IfValueTF{#2}{            {A}_{(#2)} }{            {A} } } }
\DeclareDocumentCommand{\Fg}{ s o }{ \IfBooleanTF{#1}
    { \IfValueTF{#2}{ \bm{\mathcal{F}}_{(#2)} }{ \bm{\mathcal{F}} } }
    { \IfValueTF{#2}{    {\mathcal{F}}_{(#2)} }{    {\mathcal{F}} } } }
\DeclareDocumentCommand{\Ff}{ s o }{ \IfBooleanTF{#1}
    { \IfValueTF{#2}{ \boldsymbol{F}_{(#2)} }{ \boldsymbol{F} } }
    { \IfValueTF{#2}{            {F}_{(#2)} }{            {F} } } }

\DeclareMathOperator{\st}{\star}

\DeclareDocumentCommand{\PB}{ O{m} O{q} O{p} m m }{ \frac{ \partial #4 }{\partial {#2}^{#1} } \frac{ \partial #5 }{\partial {#3}_{#1} } - \frac{ \partial #4 }{\partial {#3}_{#1} } \frac{ \partial #5 }{\partial {#2}^{#1} } }

\newcommand{\R}{\mathbb{R}}

\DeclareDocumentCommand\Te{o o m }{\mathcal{T}{}^{#1}_{#2}(#3)}

\DeclareDocumentCommand{\BM}{ s }{ \IfBooleanTF{#1} {\hat{\bm{M}}}{\bm{M}} }
\DeclareDocumentCommand{\BN}{ s }{ \IfBooleanTF{#1} {\hat{\bm{N}}}{\bm{N}} }
\DeclareDocumentCommand{\BP}{ s }{ \IfBooleanTF{#1} {\hat{\bm{P}}}{\bm{P}} }
\DeclareDocumentCommand{\BQ}{ s }{ \IfBooleanTF{#1} {\hat{\bm{Q}}}{\bm{Q}} }
\DeclareDocumentCommand{\BR}{ s }{ \IfBooleanTF{#1} {\hat{\bm{R}}}{\bm{R}} }
\DeclareDocumentCommand{\BS}{ s }{ \IfBooleanTF{#1} {\hat{\bm{S}}}{\bm{S}} }
\DeclareDocumentCommand{\BU}{ s }{ \IfBooleanTF{#1} {\hat{\bm{U}}}{\bm{U}} }
\DeclareDocumentCommand{\BV}{ s }{ \IfBooleanTF{#1} {\hat{\bm{V}}}{\bm{V}} }

\NewDocumentCommand\MyAc{ m }{#1}
\DeclareDocumentCommand{\vif}{ t. t, t- s s m }{
  \RenewDocumentCommand\MyAc{ m }{##1}
  \IfBooleanT{#1}{\RenewDocumentCommand\MyAc{ m }{ \mathring{##1} } }
  \IfBooleanT{#2}{\RenewDocumentCommand\MyAc{ m }{ \tilde{##1} } }
  \IfBooleanT{#3}{\RenewDocumentCommand\MyAc{ m }{ \bar{##1} } }
  \IfBooleanTF{#4}
  { \IfBooleanTF{#5} { \hat{\MyAc{\boldsymbol{e}}}^{\hat{#6}} }{ \hat{\MyAc{\boldsymbol{e}}}^{{#6}} } }
  { \MyAc{\boldsymbol{e}}^{{#6}} } }
\DeclareDocumentCommand{\vi}{ t. t, t- s s m m}{
  \RenewDocumentCommand\MyAc{ m }{##1}
  \IfBooleanT{#1}{\RenewDocumentCommand\MyAc{ m }{ \mathring{##1} } }
  \IfBooleanT{#2}{\RenewDocumentCommand\MyAc{ m }{ \tilde{##1} } }
  \IfBooleanT{#3}{\RenewDocumentCommand\MyAc{ m }{ \bar{##1} } }
  \IfBooleanTF{#4}
  { \IfBooleanTF{#5} { \hat{\MyAc{e}}^{\hat{#6}}_{\hat{#7}} }{ \hat{\MyAc{e}}^{#6}_{{#7}} } }
  { \MyAc{e}^{{#6}}_{{#7}} } }

\DeclareDocumentCommand{\bt}{ t. t, t- s s m m m }{
  \RenewDocumentCommand\MyAc{ m }{##1}
  \IfBooleanT{#1}{\RenewDocumentCommand\MyAc{ m }{ \mathring{##1} } }
  \IfBooleanT{#2}{\RenewDocumentCommand\MyAc{ m }{ \tilde{##1} } }
  \IfBooleanT{#3}{\RenewDocumentCommand\MyAc{ m }{ \bar{##1} } }
  \IfBooleanTF{#4}
  { \IfBooleanTF{#5} { \hat{\MyAc{B}}_{{#6}}{}^{\hat{#7}}{}_{\hat{#8}} }{ \hat{\MyAc{\Gamma}}_{{#6}}{}^{{#7}}{}_{{#8}} } }
  { \MyAc{B}_{{#6}}{}^{{#7}}{}_{{#8}} } }

\DeclareDocumentCommand{\ct}{ t. t, t- s s m m m }{
  \RenewDocumentCommand\MyAc{ m }{##1}
  \IfBooleanT{#1}{\RenewDocumentCommand\MyAc{ m }{ \mathring{##1} } }
  \IfBooleanT{#2}{\RenewDocumentCommand\MyAc{ m }{ \tilde{##1} } }
  \IfBooleanT{#3}{\RenewDocumentCommand\MyAc{ m }{ \bar{##1} } }
  \IfBooleanTF{#4}
  { \IfBooleanTF{#5} { \hat{\MyAc{\Gamma}}_{{#6}}{}^{\hat{#7}}{}_{\hat{#8}} }{ \hat{\MyAc{\Gamma}}_{{#6}}{}^{{#7}}{}_{{#8}} } }
  { \MyAc{\Gamma}_{{#6}}{}^{{#7}}{}_{{#8}} } }
\DeclareDocumentCommand{\spif}{ t. t, t- s s m m }{
  \RenewDocumentCommand\MyAc{ m }{##1}
  \IfBooleanT{#1}{\RenewDocumentCommand\MyAc{ m }{ \mathring{##1} } }
  \IfBooleanT{#2}{\RenewDocumentCommand\MyAc{ m }{ \tilde{##1} } }
  \IfBooleanT{#3}{\RenewDocumentCommand\MyAc{ m }{ \bar{##1} } }
  \IfBooleanTF{#4}
  { \IfBooleanTF{#5} { \hat{\MyAc{\boldsymbol{\omega}}}^{\hat{#6}}{}_{\hat{#7}} }{ \hat{\MyAc{\boldsymbol{\omega}}}^{{#6}}{}_{{#7}} } }
  { \MyAc{\boldsymbol{\omega}}^{{#6}}{}_{{#7}} } }
\DeclareDocumentCommand{\spi}{ t. t, t- s s m m m }{
  \RenewDocumentCommand\MyAc{ m }{##1}
  \IfBooleanT{#1}{\RenewDocumentCommand\MyAc{ m }{ \mathring{##1} } }
  \IfBooleanT{#2}{\RenewDocumentCommand\MyAc{ m }{ \tilde{##1} } }
  \IfBooleanT{#3}{\RenewDocumentCommand\MyAc{ m }{ \bar{##1} } }
  \IfBooleanTF{#4}
  { \IfBooleanTF{#5} { \hat{\MyAc{{\omega}}}_{\hat{#6}}{}^{\hat{#7}}{}_{\hat{#8}} }{ \hat{\MyAc{{\omega}}}_{{#6}}{}^{{#7}}{}_{{#8}} } }
  { \MyAc{{\omega}}_{{#6}}{}^{{#7}}{}_{{#8}} } }

\DeclareDocumentCommand{\rif}{ t. t, t- s s m m }{
  \RenewDocumentCommand\MyAc{ m }{##1}
  \IfBooleanT{#1}{\RenewDocumentCommand\MyAc{ m }{ \mathring{##1} } }
  \IfBooleanT{#2}{\RenewDocumentCommand\MyAc{ m }{ \tilde{##1} } }
  \IfBooleanT{#3}{\RenewDocumentCommand\MyAc{ m }{ \bar{##1} } }
  \IfBooleanTF{#4}
  { \IfBooleanTF{#5} { \hat{\MyAc{\bm{\mathcal{R}}}}{}^{\hat{#6}}{}_{\hat{#7}} }{ \hat{\MyAc{\bm{\mathcal{R}}}}{}^{{#6}}{}_{{#7}} } }
  { \MyAc{\bm{\mathcal{R}}}{}^{{#6}}{}_{{#7}} } }
\DeclareDocumentCommand{\ri}{ t. t, t- s s m m m }{
  \RenewDocumentCommand\MyAc{ m }{##1}
  \IfBooleanT{#1}{\RenewDocumentCommand\MyAc{ m }{ \mathring{##1} } }
  \IfBooleanT{#2}{\RenewDocumentCommand\MyAc{ m }{ \tilde{##1} } }
  \IfBooleanT{#3}{\RenewDocumentCommand\MyAc{ m }{ \bar{##1} } }
  \IfBooleanTF{#4}
  { \IfBooleanTF{#5} { \hat{\MyAc{\mathcal{R}}}_{{#6}}{}^{\hat{#7}}{}_{\hat{#8}} }{ \hat{\MyAc{\mathcal{R}}}_{{#6}}{}^{{#7}}{}_{{#8}} } }
  { \MyAc{\mathcal{R}}_{{#6}}{}^{{#7}}{}_{{#8}} } }

\newcommand{\Rif}[2]{\bm{\mathcal{R}}^{{#1}}{}_{{#2}}}

\DeclareDocumentCommand{\kf}{ t. t, t- s s m m }{
  \RenewDocumentCommand\MyAc{ m }{##1}
  \IfBooleanT{#1}{\RenewDocumentCommand\MyAc{ m }{ \mathring{##1} } }
  \IfBooleanT{#2}{\RenewDocumentCommand\MyAc{ m }{ \tilde{##1} } }
  \IfBooleanT{#3}{\RenewDocumentCommand\MyAc{ m }{ \bar{##1} } }
  \IfBooleanTF{#4}
  { \IfBooleanTF{#5} { \hat{\MyAc{\bm{\mathcal{K}}}}^{\hat{#6}}{}_{\hat{#7}} }{ \hat{\MyAc{\bm{\mathcal{K}}}}^{{#6}}{}_{{#7}} } }
  { \MyAc{\bm{\mathcal{K}}}^{{#6}}{}_{{#7}} } }
\DeclareDocumentCommand{\ko}{ t. t, t- s s m m m }{
  \RenewDocumentCommand\MyAc{ m }{##1}
  \IfBooleanT{#1}{\RenewDocumentCommand\MyAc{ m }{ \mathring{##1} } }
  \IfBooleanT{#2}{\RenewDocumentCommand\MyAc{ m }{ \tilde{##1} } }
  \IfBooleanT{#3}{\RenewDocumentCommand\MyAc{ m }{ \bar{##1} } }
  \IfBooleanTF{#4}
  { \IfBooleanTF{#5} { \hat{\MyAc{\mathcal{K}}}_{\hat{#6}}{}^{\hat{#7}}{}_{\hat{#8}} }{ \hat{\MyAc{\mathcal{K}}}_{{#6}}{}^{{#7}}{}_{{#8}} } }
  { \MyAc{\mathcal{K}}_{{#6}}{}^{{#7}}{}_{{#8}} } }

\DeclareDocumentCommand{\tf}{ t. t, t- s s m }{
  \RenewDocumentCommand\MyAc{ m }{##1}
  \IfBooleanT{#1}{\RenewDocumentCommand\MyAc{ m }{ \mathring{##1} } }
  \IfBooleanT{#2}{\RenewDocumentCommand\MyAc{ m }{ \tilde{##1} } }
  \IfBooleanT{#3}{\RenewDocumentCommand\MyAc{ m }{ \bar{##1} } }
  \IfBooleanTF{#4}
  { \IfBooleanTF{#5} { \hat{\MyAc{\bm{\mathcal{T}}}}^{\hat{#6}} }{ \hat{\MyAc{\bm{\mathcal{T}}}}^{{#6}} } }
  { \MyAc{\bm{\mathcal{T}}}^{{#6}} } }

\newcommand\miletra{}
\DeclareDocumentCommand{\PG}{ s O{\Gamma} m m m }{
  \IfBooleanTF{#1}{
    \renewcommand\miletra{#2}
  }{
    \renewcommand\miletra{\Pi_{#2}}
  }
  \miletra{}^{#3}{}_{#4}{}^{#5}
}

\newcommand{\comm}[2]{\left[#1,#2\right]}

\renewcommand{\set}[1]{\ensuremath{\Set{ #1 }}}

\newcommand{\Ric}{\operatorname{Ric}}
\newcommand*{\diag}{\operatorname{diag}}

\newcommand{\Tr}{\operatorname{Tr}}

\newcommand{\beq}{\begin{equation}}
\newcommand{\eeq}{\end{equation}}
\newcommand{\ber}{\begin{eqnarray}}
\newcommand{\eer}{\end{eqnarray}}

\NewDocumentCommand{\tak}{ s m m}{
  \IfBooleanTF{#1}{ \big( {#2} \big) \big[ {#3} \big] }
              { \big( {#2} \big] \big[ {#3} \big) }
}

\newcommand{\dn}[2]{{\mathrm{d}}^{#1}\!{#2}\;}

\newcommand*{\de}[1]{\mathop{\mathrm{d}#1}\nolimits}

\mdfdefinestyle{ibox}{
  align = center,
  linecolor=green!60!black, 
  outerlinewidth = 1pt,
  frametitle={Infobox},
  frametitlerule=true,
  frametitlebackgroundcolor=green!15,
  backgroundcolor=green!10,
  roundcorner=8pt,
  frametitlealignment=\centering,
}
\newmdenv[style=ibox]{infobox}

\mdfdefinestyle{theoremstyle}{%
  linecolor=red!40,
  linewidth=2pt,%
  frametitlerule=true,%
  frametitlebackgroundcolor=gray!20,
  innertopmargin=\topskip,
}
\mdtheorem[style=theoremstyle]{cdbexample}{Cadabra Example}

\newcommand\UTFSM{Departamento de F\'isica, Universidad T\'{e}cnica Federico Santa Mar\'\i a\\ Casilla 110-V, Valpara\'iso, Chile}
\newcommand\UTFSMmat{Departamento de Matem\'aticas, Universidad T\'{e}cnica Federico Santa Mar\'\i a, \\ Casilla 110-V, Valpara\'iso, Chile}
\newcommand\CCTVal{Centro Cient\'ifico Tecnol\'ogico de Valpara\'iso\\ Casilla 110-V, Valpara\'\i so, Chile}

\usepackage{tcolorbox}
\usepgfplotslibrary{fillbetween}
\hypersetup{colorlinks=true,allcolors=blue,linktocpage=true}
\date{}
\title{Cosmological solutions to polynomial affine gravity in the torsion-free sector}
\hypersetup{
 pdfauthor={Oscar Castillo-Felisola},
 pdftitle={Cosmological solutions to polynomial affine gravity in the torsion-free sector},
 pdfkeywords={},
 pdfsubject={},
 pdfcreator={Emacs 25.1.1 (Org mode 9.1.13)}, 
 pdflang={English}}
\begin{document}

\author[1,2]{Oscar Castillo-Felisola}
\author[1]{Jos\'e Perdiguero}
\author[3]{Oscar Orellana}
\affil[1]{\UTFSM.}
\affil[2]{\CCTVal.}
\affil[3]{\UTFSMmat.}

\maketitle
\tableofcontents

\section{Introduction}
\label{sec:intro}
All of the \emph{fundamental} physics is described by four interactions:
electromagnetic, weak, strong and gravitational. The former three are
bundled into what is known as \emph{standard model} of particle physics,
which explains very accurately the physics at very short scales. These
three interactions share common grounds, e.g. they are modelled by
connections with values in a Lie algebra, they have been successfully
quantised and renormalised, and the simplest of them---Quantum
Electrodynamics---gives the most accurate results when compared with
the experiments.

On the other hand, the model that explains gravitational interaction
(General Relativity) is a field theory for the metric, which can be
thought as a potential for the gravitational connection
\cite{einstein15_zur_allgem_relat,hilbert15_grund_physik}. Although
General Relativity is the most successful theory we have to explain
gravity
\cite{will14_confr_between_geren_relat,abbott16_obser_gravit_waves_from_binar,abbott17_gravit_waves_gamma},
it cannot be formulated as a gauge theory (in four dimensions), the
standard quantisation methods lead to inconsistencies, and it is
non-renormalizable, driving the community to believe it is an
effective theory of a yet unknown fundamental one. Within the
framework of cosmology, when one wants to conciliate both \emph{standard
models},\footnote{Beside the standard model of particles, there is a \emph{standard
model of cosmology}.} it was noticed that nearly the 95\% of the Universe does
not fit into the picture. Therefore, a (huge) piece of the puzzle is
missing called the dark sector of the Universe, composed of dark
matter and dark energy. In order to solve this problem one needs to
add \emph{new physics}, by either including extra particles (say inspired in
beyond standard model physics) or changing the gravitational
sector. The latter has inspired plenty of generalisations of General
Relativity.

Although it cannot be said that the mentioned troubles are due to the
fact that the model is described by the metric, given that the
\emph{physical} quantity associated with the gravitational
interaction---the curvature---is defined for a connection, it is worth
to ask ourselves whether a \emph{more} fundamental model of gravitational
interactions can be built up using the affine connection as the
mediator.

The first affine model of gravity was proposed by Sir A. Eddington in
Ref. \cite{eddington23}, were the action was defined by the square root
of the determinant of the Ricci tensor,
\begin{equation*}
  S = \int \sqrt{\det(\Ric)},
\end{equation*}
but in Schrödinger's words \cite{schroedinger50_space}
\begin{quote}
For all that I know, no special solution has yet been found which
suggests an application to anything that might interest us, \ldots{}
\end{quote}
However, Eddington's idea serves as starting point to new proposals
\cite{banados10_eddin_theor_gravit_its_progen,banados14_errat}.

In a series of seminal papers
\cite{cartan22_sur_une_de_la_notion,cartan23_sur_les_connex_affin_et,cartan24_sur_les_connex_affin_et,cartan25_sur_les_connex_affin_et},
E. Cartan presented a definition of curvature for spaces with torsion,
and its relevance for General Relativity. It is worth mentioning, that
in pure gravity---described by the Einstein--Hilbert-like action---, Cartan's
generalisation of gravity yields the condition of vanishing torsion as
an equation of motion. Therefore, it was not seriously considered as a
generalisation of General Relativity, until the inclusion of
gravitating fermionic matter \cite{kibble61_loren_invar_gravit}. 

Inspired by Cartan's idea of considering an affine connection into the
modelling of gravity, new interesting proposal have been
considered. Among the interesting generalisations we mention a couple:
(i) The well-known metric-affine models of gravity
\cite{hehl95_metric_affin_gauge_theor_gravit}, in which the metric and
connection are not only considered as independent, but the conditions
of metricity and vanishing torsion are in general dropped; (ii) The
Lovelock--Cartan gravity \cite{mardones91_lovel}, includes extra terms
in the action compatible with the precepts of General Relativity,
whose variation yield field equations that are second order
differential equations. Nonetheless, the metric plays a very important
role in these models.

Modern attempts to describe gravity as a theory for the affine
connection have been proposed in
Refs. \cite{kijowski78_new_variat_princ_gener_relat,krasnov07_non_metric_gravit,krasnov08_non_metric_gravit_I,krasnov11_pure_connec_action_princ_gener_relat,poplawski07_unified_purel_affin_theor_gravit_elect,poplawski07_nonsy_purel_affin,poplawski14_affin_theor,castillo-felisola15_polyn_model_purel_affin_gravit,castillo-felisola18_einst_gravit_from_polyn_affin_model},
and the cosmological implications in a
Eddington-inspired affine model were studied in Refs.
\cite{azri17_affin_inflat,azri18_induc_affin_inflat,azri18_are_there_reall_confor_frames,azri18_cosmol_implic_affin_gravit}.  

The recently proposed Polynomial Affine Gravity
\cite{castillo-felisola15_polyn_model_purel_affin_gravit}, separates the
two roles of the metric field, as in a Palatini formulation of
gravity, but does not allow it to participate in the mediation of the
interaction, by its exclusion from the action. It turns out that the
absence of the metric in the action, results in a robust structure
that---without the addition of other fields---does not accept
deformations. That robustness can be useful if one would like to
quantise the theory, because all possible counter-terms should have
the form of terms already present in the action.

In this paper, we focus in finding cosmological solutions in the
context of Polynomial Affine Gravity, restricted to torsion-free
sector of equi-affine connections, which yields a simple set of field
equations generalising those obtained in standard General Relativity
\cite{castillo-felisola18_einst_gravit_from_polyn_affin_model}. This
paper is divided into four sections: In Section \ref{sec.PAG} we review
briefly the polynomial affine model of gravity.  In Section
\ref{sec.metr_sol} we use the Levì-Civita connection for a
Friedman--Robertson--Walker metric, to solve the field
equations---obtained in the torsion-free sector---of Polynomial Affine
Gravity. Then, in Section \ref{sec.non_metr_sol} we solve for, the case
of (affine) Ricci flat manifold, the field equations for the affine
connection. Some remarks and conclusions are presented in Section
\ref{sec.concl}. In order for completeness, in Appendix \ref{sec.Lie_der},
we include a short exposition of the Lie derivative applied to the
connection, and show the Killing vectors compatible with the
cosmological principle.

\section{Polynomial Affine Gravity}
\label{sec.PAG}
In the standard theory of gravity, General Relativity, the fundamental
field is the metric, \(g_{\mu\nu}\), of the spacetime
\cite{einstein15_zur_allgem_relat,hilbert15_grund_physik}. Nevertheless,
the metric has a two fold role in this gravitational model: it
measures distances, and also define the notion of parallelism,
i.e. settles the connection. Palatini, in
Ref. \cite{palatini19_deduz_invar_delle_equaz_gravit}, considered a
somehow separation of these roles, but at the end of the day the
metric was still the sole field of the model. It was understood soon
after that the connection, \(\Gamma^\mu{}_{\rho\sigma}\), does not need
to be related with the metric field
\cite{cartan22_sur_une_de_la_notion,cartan23_sur_les_connex_affin_et,cartan24_sur_les_connex_affin_et,cartan25_sur_les_connex_affin_et,debever79_elie_cartan_alber_einst_letter},
and the therefore, the curvature could be blind to the metric.

In this section we briefly expose the model proposed in
Refs. \cite{castillo-felisola15_polyn_model_purel_affin_gravit,castillo-felisola18_einst_gravit_from_polyn_affin_model},
which is inspired in the aforementioned role separation. The metric is
left out the mediation of gravitational interactions by taking it out
the action.

The action of the polynomial affine gravity is built up from an affine
connection, \(\hat{\Gamma}^\mu{}_{\rho\sigma}\), which accepts a
decomposition on irreducible components as
\begin{equation}
\label{eq.gamm_decom}
  \hat{\Gamma}^\mu{}_{\rho\sigma} 
  = \hat{\Gamma}^\mu{}_{(\rho\sigma)} + \hat{\Gamma}^\mu{}_{[\rho\sigma]} 
  = {\Gamma}^\mu{}_{\rho\sigma} + \epsilon_{\rho\sigma\lambda\kappa}T^{\mu,\lambda\kappa}+A_{[\rho}\delta^\mu_{\nu]},
\end{equation} 
where \({\Gamma}^\mu{}_{\rho\sigma} =
\hat{\Gamma}^\mu{}_{(\rho\sigma)}\) is symmetric in the lower indices,
\(A_\rho\) is a vector field corresponding to the trace of torsion, and
\(T^{\mu,\lambda\kappa}\) is a Curtright field \cite{curtright85_gener_gauge_field},
which satisfy the properties \(T^{\kappa,\mu\nu } = -T^{\kappa,\nu\mu}\)
and \(\epsilon_{\lambda\kappa\mu\nu}T^{\kappa,\mu\nu}=0\).  The metric
field, which might or might not exists, cannot be used for contracting
nor lowering or raising indices.  The relation between the epsilons
with lower and upper indices is given by
\(\epsilon^{\delta\eta\lambda\kappa}\epsilon_{\mu\nu\rho\sigma}=4!\delta^{\delta}{}_{[\mu}\delta^\eta{}_{\nu}\delta^{\lambda}{}_{\rho}
\delta^\kappa{}_{\sigma]}\).

The most general action preserving diffeomorphisms invariance, written
in terms of the fields in Eq. \eqref{eq.gamm_decom}, is
\begin{dmath}
  \label{eq.4dfull}
  S[{\Gamma},T,A] =
  \int\dn{4}{x}\bigg[
    B_1\, R_{\mu\nu}{}^{\mu}{}_{\rho} T^{\nu,\alpha\beta}T^{\rho,\gamma\delta}\epsilon_{\alpha\beta\gamma\delta}
    +B_2\, R_{\mu\nu}{}^{\sigma}{}_\rho T^{\beta,\mu\nu}T^{\rho,\gamma\delta}\epsilon_{\sigma\beta\gamma\delta}
    +B_3\, R_{\mu\nu}{}^{\mu}{}_{\rho} T^{\nu,\rho\sigma}A_\sigma
    +B_4\, R_{\mu\nu}{}^{\sigma}{}_\rho T^{\rho,\mu\nu}A_\sigma
    +B_5\, R_{\mu\nu}{}^{\rho}{}_\rho T^{\sigma,\mu\nu}A_\sigma
    +C_1\, R_{\mu\rho}{}^{\mu}{}_\nu \nabla_\sigma T^{\nu,\rho\sigma}
    +C_2\, R_{\mu\nu}{}^{\rho}{}_\rho \nabla_\sigma T^{\sigma,\mu\nu} 
    +D_1\, T^{\alpha,\mu\nu}T^{\beta,\rho\sigma}\nabla_\gamma T^{(\lambda, \kappa) \gamma}\epsilon_{\beta\mu\nu\lambda}\epsilon_{\alpha\rho\sigma\kappa}
    +D_2\,T^{\alpha,\mu\nu}T^{\lambda,\beta\gamma}\nabla_\lambda T^{\delta,\rho\sigma}\epsilon_{\alpha\beta\gamma\delta}\epsilon_{\mu\nu\rho\sigma}
    +D_3\,T^{\mu,\alpha\beta}T^{\lambda,\nu\gamma}\nabla_\lambda T^{\delta,\rho\sigma}\epsilon_{\alpha\beta\gamma\delta}\epsilon_{\mu\nu\rho\sigma}
    +D_4\,T^{\lambda,\mu\nu}T^{\kappa,\rho\sigma}\nabla_{(\lambda} A_{\kappa)} \epsilon_{\mu\nu\rho\sigma}
    +D_5\,T^{\lambda,\mu\nu}\nabla_{[\lambda}T^{\kappa,\rho\sigma} A_{\kappa]} \epsilon_{\mu\nu\rho\sigma}
    +D_6\,T^{\lambda,\mu\nu}A_\nu\nabla_{(\lambda} A_{\mu)}
    +D_7\,T^{\lambda,\mu\nu}A_\lambda\nabla_{[\mu} A_{\nu]} 
    +E_1\,\nabla_{(\rho} T^{\rho,\mu\nu}\nabla_{\sigma)} T^{\sigma,\lambda\kappa}\epsilon_{\mu\nu\lambda\kappa}
    +E_2\,\nabla_{(\lambda} T^{\lambda,\mu\nu}\nabla_{\mu)} A_\nu
    +T^{\alpha,\beta\gamma}T^{\delta,\eta\kappa}T^{\lambda,\mu\nu}T^{\rho,\sigma\tau}
    \Big(F_1\,\epsilon_{\beta\gamma\eta\kappa}\epsilon_{\alpha\rho\mu\nu}\epsilon_{\delta\lambda\sigma\tau}
    +F_2\,\epsilon_{\beta\lambda\eta\kappa}\epsilon_{\gamma\rho\mu\nu}\epsilon_{\alpha\delta\sigma\tau}\Big) 
    +F_3\, T^{\rho,\alpha\beta}T^{\gamma,\mu\nu}T^{\lambda,\sigma\tau}A_\tau \epsilon_{\alpha\beta\gamma\lambda}\epsilon_{\mu\nu\rho\sigma}
    +F_4\,T^{\eta,\alpha\beta}T^{\kappa,\gamma\delta}A_\eta A_\kappa\epsilon_{\alpha\beta\gamma\delta}\bigg],
\end{dmath}
where terms related through partial integration, and topological
invariant have been dropped.\footnote{An example of four-dimensional topological term is the Euler density.} One can prove via a dimensional
analysis, the \emph{uniqueness} of the above action (see
Ref. \cite{castillo-felisola18_einst_gravit_from_polyn_affin_model}).

The action in Eq. \eqref{eq.4dfull} shows up very interesting features:
(i) it is power-counting renormalizable,\footnote{power-counting renormalizability does not guarantee
renormalizability.} (ii) all coupling
constants are dimensionless which hints the conformal invariance of
the model \cite{buchholz77_dilat_inter}, (iii) yields no three-point
graviton vertices, which might allow to overcome the \emph{no-go} theorems
found in
Refs. \cite{mcgady14_higher_spin_massl_s_dimen,camanho16_causal_const_correc_to_gravit},
(iv) its non-relativistic geodesic deviation agrees with that produced
by a Keplerian potential
\cite{castillo-felisola15_polyn_model_purel_affin_gravit}, and (v) the
effective equations of motion in the torsion-free limit are a
generalization of the Einstein's equations
\cite{castillo-felisola18_einst_gravit_from_polyn_affin_model}. In the
remaining of this section we will sketch how to find the relativistic
limit of this model, when the torsion vanishes.

First, notice that the vanishing torsion condition is equivalent to
setting both \(T^{\lambda,\mu\nu}\) and \(A_\mu\) equal to zero. Although
this limit is not well-defined at the action level, it is well-defined
at the level of equation of motion.\footnote{The field equations can be consistently truncated under the
requirement of vanishing torsion. It is worth noticing that this
condition does not yield the Riemannian theory, since we are not yet
asking for a metricity condition.} In order to simplify the
task of finding the equations of motion to take the limit, we restrict
ourselves to the terms in the action which are linear in either
\(T^{\lambda,\mu\nu}\) or \(A_\mu\), since these are the only terms which,
after the extremisation, will survive the torsion-free
limit. Therefore, after the described considerations, the effective
torsion-free action is
\begin{equation}
  \label{eq.eff_action}
  S_{\text{eff}} = \int\dn{4}{x} \Big( C_1\, R_{\lambda\mu}{}^{\lambda}{}_\nu \nabla_\rho 
  + C_2 \, R_{\mu\rho}{}^{\lambda}{}_\lambda \nabla_\nu \Big) T^{\nu,\mu\rho}.
\end{equation}
The nontrivial equations of motion for this action are those for the
Curtright field, \(T^{\nu,\mu\rho}\),
\begin{equation}
  \nabla_{[\rho} R_{\mu]\nu} + \kappa \nabla_{\nu} R_{\mu\rho}{}^\lambda{}_\lambda = 0,
  \label{eq.almostSimpleEOM}
\end{equation}
where \(\kappa\) is a constant related with the original couplings of the
model.

In the Riemannian formulation of differential geometry, since the curvature
tersor is anti-symmetric in the last couple of indices, the second
term in Eq. \eqref{eq.almostSimpleEOM} vanishes identically. However, for
non-Riemannian connections, such term still vanishes if the connection
is compatible with a volume form. These connections are known as
equi-affine connections \cite{nomizu94_affin,bryant_symmet_rieman}. In
addition, the Ricci tensor for equi-affine connections is
symmetric. For these connections, the gravitational equations are simply
\begin{equation}
  \nabla_{[\rho} R_{\mu]\nu} = 0.
  \label{eq.SimpleEOM}
\end{equation}

The Eq. \eqref{eq.SimpleEOM} is a generalisation of the parallel Ricci
curvature condition, \(\nabla_{\rho} R_{\mu\nu} = 0\), which is a known
extension of the Einstein's equations
\cite{derdzinski80_class_certain_compac_rieman_manif,besse07_einst}. Moreover,
these field equations are also obtained as part of a à la Palatini
approach to a Yang--Mills formulation of gravity, known as the
Stephenson--Kilmister--Yang (or SKY) model, proposed in
Refs. \cite{stephenson58_quadr_lagran_gener_relat,kilmister61,yang74_integ_formal_gauge_field}. Such
Yang--Mills-like gravity is described by the action
\begin{equation}
  \label{eq.gr_YM}
  S_{\textsc{sky}} = \int \dn{4}{x} \sqrt{g} \, g^{\mu\nu} g^{\sigma\tau} \, R_{\mu\sigma}{}^\lambda{}_\rho R_{\nu\tau}{}^\rho{}_\lambda,
\end{equation}
which can be written using the curvature two-form as
\begin{equation}
  \label{eq.SKY}
  S_{\textsc{sky}} = \int \Tr \left( \Rif{}{} \st \Rif{}{} \right) = \int \left( \Rif{a}{b} \st \Rif{b}{a} \right).
\end{equation}
Although the field equations of the connection obtained from Eq. \eqref{eq.gr_YM} are the harmonic
curvature condition \cite{bourguignon81_les_de_dimen_signat_non},
\begin{equation}
  \label{eq.harm_curv}
  \nabla_\lambda R_{\mu\nu}{}^\lambda{}_\rho = 0,
\end{equation}
these are equivalent to Eq. \eqref{eq.SimpleEOM} through the second Bianchi
identity \cite{derdzinski85_rieman,besse07_einst}.

The Stephenson--Kilmister--Yang model is a field theory for the
metric---not for the connection---, and thus there is an extra field
equation for the metric. The field equation for the metric is very
restrictive, and it does not accept Schwarzschild-like solutions
\cite{zanelli_privat}. However, in the Polynomial Affine Gravity, since the
metric does not participate in the mediation of gravitational
interaction, that problem is solved trivially. Meanwhile, the
physical field associated with the gravitational interaction is the
connection. This difference makes a huge distinction in the
phenomenological interpretation of these models.

In the following sections we shall present solutions to the field
equations \eqref{eq.SimpleEOM}, in the cases where the connection is
metric or not. To this end, in appendix \ref{sec.Lie_der} we show how to
propose an ansatz compatible with the desired symmetries. Moreover,
equation  \eqref{eq.SimpleEOM} can be solved in three ways, yielding to
a sub-classification of the solutions: (i) Ricci flat solutions,
\(R_{\mu\nu} = 0\); (ii) Parallel Ricci solutions, \(\nabla_\lambda
R_{\mu\nu} = 0\); and (iii) Harmonic Riemann solutions,
\(\nabla_\lambda R_{\mu\nu}{}^\lambda{}_\rho = 0\).

\section{Cosmological metric solutions}
\label{sec.metr_sol}
The conditions of isotropy and homogeneity are very stringent, when
imposed on a symmetric rank-two tensor, and the possible ansatz is
just the Friedmann--Robertson--Walker metric,
\begin{equation}
  \label{eq.ht_metr}
  g = G_{00}\left(t\right) \mathrm{d} t\otimes \mathrm{d} t +
  G_{11}\left(t\right) \left(
  \frac{1}{1 - {\kappa} r^{2}}  \mathrm{d}
  r\otimes \mathrm{d} r + r^{2} \mathrm{d}
  {\theta}\otimes \mathrm{d} {\theta} + r^{2} 
  \sin\left({\theta}\right)^{2} \mathrm{d} {\phi}\otimes \mathrm{d} {\phi} \right).
\end{equation}

In the remaining of this section, we shall use the standard parametrisation of a
Friedmann--Robertson--Walker metric, i.e.,
\begin{equation}
  \label{eq.FRW_metr}
  g = - \mathrm{d} t\otimes \mathrm{d} t +
  a^2\left(t\right) \left(
  \frac{1}{1 - {\kappa} r^{2}}  \mathrm{d}
  r\otimes \mathrm{d} r + r^{2} \mathrm{d}
  {\theta}\otimes \mathrm{d} {\theta} + r^{2} 
  \sin\left({\theta}\right)^{2} \mathrm{d} {\phi}\otimes \mathrm{d} {\phi} \right).
\end{equation}
The nonvanishing component of the Levi-Cività connection for the
metric in Eq. \eqref{eq.FRW_metr} are
\begin{equation}
  \begin{aligned}
    \Gamma_{ \phantom{\, t} \, r \, r }^{ \, t \phantom{\, r} \phantom{\, r} } & = -\frac{a \dot{a}}{k r^{2} - 1} 
    &
    \Gamma_{ \phantom{\, t} \, {\theta} \, {\theta} }^{ \, t \phantom{\, {\theta}} \phantom{\, {\theta}} } & = r^{2} a \dot{a} 
    &
    \Gamma_{ \phantom{\, t} \, {\phi} \, {\phi} }^{ \, t \phantom{\, {\phi}} \phantom{\, {\phi}} } & = r^{2} a
    \sin^{2}\left({\theta}\right) \dot{a} 
    \\
    \Gamma_{ \phantom{\, r} \, t \, r }^{ \, r \phantom{\, t} \phantom{\, r} } & = \frac{\dot{a}}{a} 
    &
    \Gamma_{ \phantom{\, r} \, r \, t }^{ \, r \phantom{\, r} \phantom{\, t} } & = \frac{\dot{a}}{a} 
    &
    \Gamma_{ \phantom{\, r} \, r \, r }^{ \, r \phantom{\, r} \phantom{\, r} } & = -\frac{k r}{k r^{2} - 1} 
    \\
    \Gamma_{ \phantom{\, r} \, {\theta} \, {\theta} }^{ \, r \phantom{\, {\theta}} \phantom{\, {\theta}} } & = k r^{3} - r 
    &
    \Gamma_{ \phantom{\, r} \, {\phi} \, {\phi} }^{ \, r \phantom{\, {\phi}} \phantom{\, {\phi}} } & = {\left(k r^{3} - r\right)} \sin^{2}\left({\theta}\right)
    &
    \Gamma_{ \phantom{\, {\theta}} \, t \, {\theta} }^{ \, {\theta} \phantom{\, t} \phantom{\, {\theta}} } & = \frac{\dot{a}}{a} 
    \\
    \Gamma_{ \phantom{\, {\theta}} \, r \, {\theta} }^{ \, {\theta} \phantom{\, r} \phantom{\, {\theta}} } & = \frac{1}{r} 
    &
    \Gamma_{ \phantom{\, {\theta}} \, {\theta} \, t }^{ \, {\theta} \phantom{\, {\theta}} \phantom{\, t} } & = \frac{\dot{a}}{a} 
    &
    \Gamma_{ \phantom{\, {\theta}} \, {\theta} \, r }^{ \, {\theta}
      \phantom{\, {\theta}} \phantom{\, r} } & = \frac{1}{r} 
    \\
    \Gamma_{ \phantom{\, {\theta}} \, {\phi} \, {\phi} }^{ \, {\theta} \phantom{\, {\phi}} \phantom{\, {\phi}} } & = -\cos\left({\theta}\right) \sin\left({\theta}\right) 
    &
    \Gamma_{ \phantom{\, {\phi}} \, t \, {\phi} }^{ \, {\phi} \phantom{\, t} \phantom{\, {\phi}} } & = \frac{\dot{a}}{a} 
    &
    \Gamma_{ \phantom{\, {\phi}} \, r \, {\phi} }^{ \, {\phi} \phantom{\, r} \phantom{\, {\phi}} } & = \frac{1}{r} 
    \\
    \Gamma_{ \phantom{\, {\phi}} \, {\theta} \, {\phi} }^{ \, {\phi} \phantom{\, {\theta}} \phantom{\, {\phi}} } & =
    \frac{\cos\left({\theta}\right)}{\sin\left({\theta}\right)} 
    &
    \Gamma_{ \phantom{\, {\phi}} \, {\phi} \, t }^{ \, {\phi} \phantom{\, {\phi}} \phantom{\, t} } & = \frac{\dot{a}}{a} 
    &
    \Gamma_{ \phantom{\, {\phi}} \, {\phi} \, r }^{ \, {\phi} \phantom{\, {\phi}} \phantom{\, r} } & = \frac{1}{r} 
    \\
    & &
    \Gamma_{ \phantom{\, {\phi}} \, {\phi} \, {\theta} }^{ \, {\phi}
      \phantom{\, {\phi}} \phantom{\, {\theta}} } & =
    \frac{\cos\left({\theta}\right)}{\sin\left({\theta}\right)}
    & &
  \end{aligned}
  \label{eq.metr_cosmo_conn}
\end{equation}

\subsection{\ldots{} with vanishing Ricci}
\label{sec:org00c0bac}
This particular case is a metric model of gravity, whose field
equations are vanishing Ricci. It is expected to obtain the
cosmological vacuum solution of General Relativity (without
cosmological constant), i.e. Minkowski spacetime.

From the connection in Eq. \eqref{eq.metr_cosmo_conn}, it is
straightforward to calculate the Ricci tensor, and the field equations
are then
\begin{align}
  R_{tt} & =  - \frac{3 \ddot{a}}{a} = 0,
  &
  R_{ii} & = f_{i}(r,\theta) \left( \dot{a}^2 + a \ddot{a} + 2 \kappa \right) = 0,
  \label{eq.eom_metr_Ricci_flat}
\end{align}
where the functions \(f_i\) are \(f_r = (1 - \kappa r^2)^{-1}\),
\(f_\theta = r^2\) and \(f_\varphi = r^2 \sin^2 (\theta)\).

The solutions to Eqs.  \eqref{eq.eom_metr_Ricci_flat} are shown in Table
\ref{tab:metr_a_Ricci_flat}, and (as expected) are two parametrisations
of Minkowski spacetime, see for example Ref. \cite{dray14_differ}.

\tcbset{
  fonttitle=\bfseries,nobeforeafter,center title}
\begin{table}
  \begin{center}
    \tcbox[left=0mm,right=0mm,top=0mm,bottom=0mm,boxsep=0mm,
    toptitle=0.5mm,bottomtitle=0.5mm,title=Scale factor for the metric vanishing Ricci case]{%
      \begin{tabular}{p{4cm}|p{4cm}|p{4cm}}
        \(\kappa = -1\) & \(\kappa = 0\) & \(\kappa = 1\)
        \\
        \hline
        \(\sqrt{2} t + B\) & \(B \in \R^+ \) & \(\not\exists\)
      \end{tabular}
    }
    \caption{Scale factor solving the vanishing Ricci condition, for
      a cosmological metric connection}
    \label{tab:metr_a_Ricci_flat}
  \end{center}
\end{table}

\subsection{\ldots{} with parallel Ricci}
\label{sec.ht_metr_sol}
Secondly, we shall analyse the possible solutions to the parallel Ricci
equations,
\begin{equation}
  \nabla_\lambda R_{\mu\nu} = 0.
  \label{eq.parall_Ricci}
\end{equation}
Notice that in the case of Riemannian geometry, there is a \emph{natural}
parallel symmetric \(\binom{0}{2}\)-type tensor, i.e. the
metric. Therefore, a simple solution to Eq. \eqref{eq.parall_Ricci} is
that the Ricci is proportional to the metric---the spacetime is
an Einstein manifold---, and the proportionality factor is related
with the cosmological constant.

The independent components of Eq. \eqref{eq.parall_Ricci} for the
ansätze in Eq. \eqref{eq.FRW_metr} are,
\begin{align}
  \nabla_t R_{tt} & \simeq \dot{a} \ddot{a} - a \dddot{a} = 0, 
  \label{eq.oem_mpr_cosm_ttt}
  \\
  \nabla_i R_{ti} & \simeq \left( \dot{a}^2 - {a} \ddot{a} + \kappa \right) \dot{a} = 0.
  \label{eq.eom_mpr_cosm_rrt}
\end{align}
Additionally, Eq.  \eqref{eq.oem_mpr_cosm_ttt} can be rewritten as
\begin{equation}
  \label{eq.mpr_cosm_ttt}
  \frac{d}{dt} \left( \frac{\ddot{a}}{a} \right) = 0 \quad \Rightarrow \quad \ddot{a} + C a = 0.
\end{equation}
According to the value of the integration constant \(C\), we
parametrise it as
\begin{equation*}
  C = 
  \begin{cases}
    \omega^2 & \text{ for } C>0
    \\
    \omega=0 & \text{ for } C=0
    \\
    - \omega^2 & \text{ for } C<0 
  \end{cases}
\end{equation*}

Using Eq. \eqref{eq.mpr_cosm_ttt} to eliminate the \(\ddot{a}\)
dependence from Eq.  \eqref{eq.eom_mpr_cosm_rrt} yields
\begin{equation}
  \dot{a}^2 + C a^2 + \kappa = 0.
  \label{eq.mpr_cosm_rrt}
\end{equation}

\tcbset{
  fonttitle=\bfseries,nobeforeafter,center title}
\begin{table}
  \begin{center}
    \tcbox[left=0mm,right=0mm,top=0mm,bottom=0mm,boxsep=0mm,
    toptitle=0.5mm,bottomtitle=0.5mm,title=Scale factor for the metric parallel Ricci case]{%
      \begin{tabular}{p{3cm}|p{3cm}|p{3cm}|p{3cm}}
        & \(\kappa = -1\)   &   \(\kappa = 0\)   &   \(\kappa = 1\)
        \\
        \hline
        \(C = - \omega^2 < 0 \)
        & \(\pm \frac{\sinh(\omega t)}{2\omega}\)
                            & \(A \exp(\pm \omega t)\)
                                                 & \(\pm \frac{\cosh(\omega t)}{2\omega}\)
        \\
        \(C = \omega = 0\)
        & \(\pm t + B\)
                            & \(B \in \R^+ \)
                                                 & \(\not\exists\)
        \\
        \(C = \omega^2 >0\)
        & \(\frac{\sin(\omega t + \varphi)}{\omega}\)
                            & \(\not\exists\)
                                                 & \(\not\exists\)
      \end{tabular}
    }
    \caption{Scale factor solving the parallel Ricci condition, for
      a cosmological metric connection}
    \label{tab:metr_a_parallel_Ricci}
  \end{center}
\end{table}

The solutions to Eq. \eqref{eq.mpr_cosm_ttt} are presented in
Table \ref{tab:metr_a_parallel_Ricci}, and they are known from General Relativity, see
for example Ref. \cite{dray14_differ}. Interestingly our integration constant, \(C\), could be identified as \(C = -
\frac{\Lambda}{3}\) from the vacuum Friedmann's equations. However, our
equations are compatible with Friedmann's equations, interacting with
a vacuum energy perfect fluid, if the integration constant is identified with 
\begin{equation}
  C = \frac{4 \pi G_N}{3} ( \rho + 3 p ) - \frac{\Lambda}{3}.
  \label{eq.mpr_cosm_id_C}
\end{equation}

\subsection{\ldots{} with harmonic Riemann}
\label{sec:org3551bbc}
Now that we showed that the solutions of the parallel Ricci equations
are equivalent to those of General Relativity, we turn our attention
to the Eq. \eqref{eq.SimpleEOM}. For the metric ansatz in
Eq. \eqref{eq.FRW_metr}, interestingly, only an independent equation is obtained,
\begin{equation*}
  \frac{2 \dot{a}^3 - a \dot{a} \ddot{a} - a^2 \dddot{a} + 2 \kappa
    \dot{a}}{a} = 0,
\end{equation*}
that should determine the scale factor. It can be rewritten as
\begin{equation}
  - \frac{d}{dt} \left( \frac{\ddot{a}}{a} + \frac{\dot{a}^2}{a^2} + \frac{\kappa}{a^2} \right) = 0,
\end{equation}
i.e.
\begin{equation}
  \frac{\ddot{a}}{a} + \frac{\dot{a}^2}{a^2} + \frac{\kappa}{a^2}  = - C.
  \label{eq.hrim_cosm}
\end{equation}
After a change of variable, \(f = a^2\), Eq. \eqref{eq.hrim_cosm}
becomes
\begin{equation}
  \label{eq.hrim_cosm_f}
  \ddot{f} + 2 C f + 2 \kappa= 0,
\end{equation}
whose solutions are
\begin{equation}
  f(t) = 
  \begin{cases}
    - \kappa t^2 + A t + B & C = 0
    \\
    A \sin( \omega t ) + B \cos( \omega t ) - \frac{2 \kappa}{\omega^2} & 2 C = \omega^2 > 0
    \\
    A \sinh( \omega t ) + B \cosh( \omega t ) + \frac{2 \kappa}{\omega^2} & 2 C = - \omega^2 <0
  \end{cases}
\end{equation}
Therefore, the scale factors are those presented in Table
\ref{tab:metr_a_harm_curv}. Notice, however, that in this case we are
not separating  the cases according to the value of \(\kappa\), but
the existence of a solution for a given \(\kappa\) is determined by
the domain of time, and also by the values of the integration
constants \(A\) and \(B\).
\tcbset{
  fonttitle=\bfseries,nobeforeafter,center title}
\begin{table}
  \begin{center}
    \tcbox[left=0mm,right=0mm,top=0mm,bottom=0mm,boxsep=0mm,
    toptitle=0.5mm,bottomtitle=0.5mm,title=Scale factor for the metric harmonic curvature case]{%
      \begin{tabular}{p{3cm}|>{\hfil}p{12cm}<{\hfil}}
        & \(\kappa = {-1,0,1}\)   
        \\
        \hline
        \(C = - \omega^2 < 0 \)
        & \( \sqrt{ A \sinh(\omega t) + B \cosh(\omega t) + \frac{2 \kappa}{\omega^2} }\)
        \\
        \(C = \omega = 0\)
        & \( \sqrt{ - \kappa t^2 + A t + B } \)
        \\
        \(C = \omega^2 >0\)
        & \( \sqrt{ \frac{\sin(\omega t + \varphi)}{\omega} }\)
      \end{tabular}
    }
    \caption{Scale factor solving the harmonic curvature condition, for
      a cosmological metric connection}
    \label{tab:metr_a_harm_curv}
  \end{center}
\end{table}

\section{Cosmological non-metric solutions}
\label{sec.non_metr_sol}
In order to solve the set of coupled, non-linear, partial differential
equations for the connection, one proceeds---just as in General
Relativity---by giving an ansatz compatible with the symmetries of the
problem. Using the Lie derivative, we have found the most general
torsion-free connection compatible with the cosmological principle
\cite{castillo-felisola18_beyond_einstein}.\footnote{See appendix  \ref{sec.Lie_der} for a brief comment about the Lie
derivative of a connection.} The nonvanishing
components of the connection are,
\begin{equation}
  \begin{aligned}
    \Gamma_{ \phantom{  t }   t   t }^{   t \phantom{  t } \phantom{  t } } & = f\left(t\right) 
    &
    \Gamma_{ \phantom{  t }   r   r }^{   t \phantom{  r } \phantom{  r } } & = \frac{g\left(t\right)}{ 1 - {\kappa} r^{2} } 
    &
    \Gamma_{ \phantom{  t }   {\theta}   {\theta} }^{   t \phantom{  {\theta} } \phantom{  {\theta} } } & = r^{2} g\left(t\right) 
    \\
    \Gamma_{ \phantom{  t }   {\phi}   {\phi} }^{   t \phantom{  {\phi} } \phantom{  {\phi} } } & = r^{2} g\left(t\right) \sin^2\left({\theta}\right) 
    &
    \Gamma_{ \phantom{  r }   t   r }^{   r \phantom{  t } \phantom{  r } } & = h\left(t\right) 
    &
    \Gamma_{ \phantom{  r }   r   t }^{   r \phantom{  r } \phantom{  t } } & = h\left(t\right) 
    \\
    \Gamma_{ \phantom{  r }   r   r }^{   r \phantom{  r } \phantom{  r } } & = \frac{{\kappa} r}{ 1 - {\kappa} r^{2} } 
    &
    \Gamma_{ \phantom{  r }   {\theta}   {\theta} }^{   r \phantom{  {\theta} } \phantom{  {\theta} } } & = {\kappa} r^{3} - r 
    &
    \Gamma_{ \phantom{  r }   {\phi}   {\phi} }^{   r \phantom{  {\phi} } \phantom{  {\phi} } } & = {\left({\kappa} r^{3} - r\right)} \sin^2\left({\theta}\right) 
    \\
    \Gamma_{ \phantom{  {\theta} }   t   {\theta} }^{   {\theta} \phantom{  t } \phantom{  {\theta} } } & = h\left(t\right) 
    &
    \Gamma_{ \phantom{  {\theta} }   r   {\theta} }^{   {\theta} \phantom{  r } \phantom{  {\theta} } } & = \frac{1}{r} 
    &
    \Gamma_{ \phantom{  {\theta} }   {\theta}   t }^{   {\theta} \phantom{  {\theta} } \phantom{  t } } & = h\left(t\right) 
    \\
    \Gamma_{ \phantom{  {\theta} }   {\theta}   r }^{   {\theta} \phantom{  {\theta} } \phantom{  r } } & = \frac{1}{r} 
    &
    \Gamma_{ \phantom{  {\theta} }   {\phi}   {\phi} }^{   {\theta} \phantom{  {\phi} } \phantom{  {\phi} } } & = -\cos\left({\theta}\right) \sin\left({\theta}\right) 
    &
    \Gamma_{ \phantom{  {\phi} }   t   {\phi} }^{   {\phi} \phantom{  t } \phantom{  {\phi} } } & = h\left(t\right) 
    \\
    \Gamma_{ \phantom{  {\phi} }   r   {\phi} }^{   {\phi} \phantom{  r } \phantom{  {\phi} } } & = \frac{1}{r} 
    &
    \Gamma_{ \phantom{  {\phi} }   {\theta}   {\phi} }^{   {\phi} \phantom{  {\theta} } \phantom{  {\phi} } } & = \frac{\cos\left({\theta}\right)}{\sin\left({\theta}\right)} 
    &
    \Gamma_{ \phantom{  {\phi} }   {\phi}   t }^{   {\phi} \phantom{  {\phi} } \phantom{  t } } & = h\left(t\right) 
    \\
    \Gamma_{ \phantom{  {\phi} }   {\phi}   r }^{   {\phi} \phantom{  {\phi} } \phantom{  r } } & = \frac{1}{r} 
    &
    \Gamma_{ \phantom{  {\phi} }   {\phi}   {\theta} }^{   {\phi} \phantom{  {\phi} } \phantom{  {\theta} } } & = \frac{\cos\left({\theta}\right)}{\sin\left({\theta}\right)}
    &
  \end{aligned}
  \label{eq.ht_conn_torsion_free}
\end{equation}
with \(f\), \(g\) and \(h\) the unknown functions of time to be
determined. The Levi-Cività connection compatible with the
Friedman--Robertson--Walker metric is obtained from
Eq. \eqref{eq.ht_conn_torsion_free} by setting \(f = 0\), \(g = a
\dot{a}\) and \(h = \frac{\dot{a}}{a}\)---Compare with Eq. \eqref{eq.metr_cosmo_conn}.

The Ricci tensor calculated for the connection in
Eq. \eqref{eq.ht_conn_torsion_free} has only two independent components
\begin{align}
  R_{  t  t }
  & =  3  f h - 3 
  h^{2} - 3  \frac{\partial h}{\partial t}, 
  \\
  R_{  i i }
  & \simeq 
  {f g +
    g h + 2  {\kappa} +
    \frac{\partial g}{\partial t}}.
\end{align}

We now proceed to find solutions to Eq.  \eqref{eq.SimpleEOM}. As in the
previous section, we present the three possibilities of solutions, but
we will restrict ourselves to finding solutions to the (affine) Ricci
flat case.

\subsection{\ldots{} with vanishing Ricci}
\label{sec:org5aa5ec5}
A first kind of solutions can be found by solving the system of
equations determined by vanishing Ricci. However, this strategy
requires the fixing of one of the unknown functions. The equations to
solve are written as
\begin{align}
  \dot{h} - (f - h) h & = 0,
  \label{eq.affine_cosm_rf_1}
  \\
  \dot{g} + (f + h) g + 2 \kappa & = 0. 
  \label{eq.affine_cosm_rf_2}
\end{align}

Noticing that in the above equations \(f\) is not a dynamical
function, from Eq. \eqref{eq.affine_cosm_rf_1} we can solve \(h\)
as a function of \(f\),
\begin{equation}
  h(t) = \frac{\exp \left( F(t) \right)}{C_h + \int \de{t} \,\exp(F)},
  \label{eq.aff_cosm_fr_sol_h}
\end{equation}
where we have defined \(F = \int \de{t} \, f(t)\) and \(C_h\) is an
integration constant. Then, Eq.  \eqref{eq.affine_cosm_rf_2} can be
solved for \(g\),
\begin{equation}
  g(t) = \exp( - \Sigma(t) ) \left( C_g - 2 \kappa \left( \int \de{t} \exp( \Sigma(t) )
  \right) \right),
  \label{eq.aff_cosm_fr_sol_g}
\end{equation}
where \(\Sigma(t) = \int \de{t} \, \left( f(t) + h(t) \right)\), and
\(C_g\) is another integration constant.

A particular solution inspired in the components of the connection for
Friedmann--Robertson--Walker, in whose case \(f = 0\), gives
\begin{equation}
  \begin{aligned}
    f(t) &= 0,
    &
    g(t) &= \frac{1}{t + C_h} \left( C_g - \kappa (t + C_h)^2 \right) ,
    &
    h(t) &= \frac{1}{t + C_h},
  \end{aligned}
  \label{eq.sol_cosm_very_simple}
\end{equation}
which for \(C_h = C_g = 0\) and \(\kappa = -1\) yields the expected
solution from Table \ref{tab:metr_a_Ricci_flat}.\footnote{The standard parametrisation of Minkowski spacetime is achived
by the trivial solution of Eqs. \eqref{eq.affine_cosm_rf_1} and \eqref{eq.affine_cosm_rf_2}, i.e.,
\(f = g = h = \kappa = 0\).} However, in
Eq. \eqref{eq.sol_cosm_very_simple} there are Ricci flat solutions which
cannot be associated with the sole existence of a metric,
i.e. non-Riemannian manifolds; as for example solutions with \(\kappa
> 0\).

There are special solutions that cannot be obtained from Eqs.
\eqref{eq.aff_cosm_fr_sol_h} and \eqref{eq.aff_cosm_fr_sol_g}, since they
represent degenerated point in the moduli space.

\begin{description}
\item[{Case \(f = h\):}] In these particular subspace on the Moduli, the first equation is
linear, and therefore the solution above is not valid. However, the
solutions to Eqs. \eqref{eq.affine_cosm_rf_1} and
\eqref{eq.affine_cosm_rf_2} are given by
\begin{align}
  f   & = C_h
  &
    h & = C_h
  & g & = C_g \exp( - 2 C t ) - \frac{\kappa}{C_h}.
\end{align}
\end{description}

\begin{description}
\item[{Case \(h = - f\):}] In this case again Eq. \eqref{eq.affine_cosm_rf_1} decouples from
Eq. \eqref{eq.affine_cosm_rf_2}, and there solutions are given by
\begin{align}
  f   & =  - \frac{1}{2t + C_h}
  &
    h & =  \frac{1}{2t + C_h}
  &
    g & = C_g - 2 \kappa t
\end{align}

\item[{Case \(h = 0\) and \(f\) given:}] In this case, Eq.  \eqref{eq.affine_cosm_rf_1} becomes an identity, and
\(g\) can still be solved for a given function \(f\) as
\begin{equation}
  g(t) = \exp( - F(t) ) \left( C_g - 2 \kappa \left( \int \de{t} \exp( F(t) )
  \right) \right).
\end{equation}

\item[{Case \(g = 0\) and \(f\) given:}] In this case, Eq.  \eqref{eq.affine_cosm_rf_2} requires \(\kappa = 0\),
and \(h\) can still be solved for a given function \(f\) as in Eq.
\eqref{eq.aff_cosm_fr_sol_h}.
\end{description}

At this point, we have shown that a spacetime described by a Ricci
flat, torsion-free, equi-affine connection with the form presented in
Eq. \eqref{eq.ht_conn_torsion_free} reproduces the cosmological Ricci
flat solutions to General Relativity, presented in Table
\ref{tab:metr_a_Ricci_flat}, and there exist generalisations to these
solutions which are not possibly obtained in the Riemannian
case. However, one can go even further, and ask oneself whether the
affine Ricci flat condition yield more---real life---useful solutions,
such as those solutions of General Relativity presented in Table
\ref{tab:metr_a_parallel_Ricci}.

Therefore, we would like to obtain the Einstein equations from the
affine Ricci flat equation, i.e.,
\begin{equation}
  \label{eq:aff_Ricci_flat_vs_GR}
  R^{\text{Aff}}_{\mu\nu} = M^{\text{GR}}_{\mu\nu} = 0,
\end{equation}
where
\begin{equation*}
  M^{\text{GR}}_{\mu\nu} = R^{\text{GR}}_{\mu\nu} - \Lambda g_{\mu\nu} - 8 \pi G  \left( T_{\mu\nu} - \frac{1}{2} T g_{\mu\nu}\right).
\end{equation*}
In the following, we are considering that the stress-energy tensor
describes a perfect fluid, i.e.,  
\[T_{\mu\nu} = \diag \begin{pmatrix} \rho & \frac{p a^2}{1 - \kappa r^2} & p a^2 r^2 & p a^2 r^2 \sin^2 \theta \end{pmatrix}.\]

In General Relativity, the Einstein equations in the form of Ricci,
for the cosmological ansatz yields
\begin{align}
  M^{\text{GR}}_{tt} & \simeq 3\ddot{a} - \Lambda a + 4 \pi G a (
  \rho + 3 p ),
  &
    M^{\text{GR}}_{tt} & \simeq a \ddot{a} + 2 \dot{a}^2 - \Lambda a^2
                         - 4 \pi G a^2 ( \rho - p ) + 2 \kappa,
  \label{eq.cosmo_Eins_eq_Ricci_form_GR}
\end{align}
Now, comparing Eq. \eqref{eq.cosmo_Eins_eq_Ricci_form_GR} with
Eqs. \eqref{eq.affine_cosm_rf_1}  and \eqref{eq.affine_cosm_rf_2}, a
parametrisation for \(f\), \(g\) and \(h\) can be found such that once
one compute the Ricci tensor for the affine connection, the
compatibility in Eq. \eqref{eq:aff_Ricci_flat_vs_GR} is satisfied. The
parametrisation is given by
\begin{align}
  \label{eq.fgh_param_pr}
  h & = \dot{a} + x, & f & = x, & g & = a \dot{a} + y,
\end{align}
where the functions \(x\) and \(y\) satisfy the equations,
\begin{align}
  \label{eq:ODE_x_func}
  \dot{x}  + x \dot{a} - F_1 & = 0,
  \\
  \label{eq:ODE_y_func}
  \dot{y}  + y F_2 - F_3 & = 0,
\end{align}
with
\begin{align}
  F_1 & = \frac{a}{3}\Big(4\pi G\left(3p + \rho\right) -\Lambda\Big) - (\dot{a})^2, &  F_2 & = \dot{a} +2 x, & F_3 & = a^{2}\left(4\pi G (p - \rho) - \Lambda \right) -2a\dot a x - a(\dot a)^2 + (\dot a)^2.
\end{align}

The Eqs. \eqref{eq:ODE_x_func} and \eqref{eq:ODE_y_func} can be formally
integrated in terms of functions \(a\), \(\rho\) and \(p\), yielding 
\begin{align}
  x & = e^{-a} \left( C_x + \int \de{t}  F_1 \, e^a \right),
      \label{eq.sol_x_func}
  \\
  y & = e^{- \int \de{t} F_2} \left( C_y + \int \de{t}  F_3 \, e^{\int \de{t} F_2} \right).
      \label{eq.sol_y_func}
\end{align}
Therefore, a subspace of the possible solutions of the affine Ricci
flat geometries, describes the cosmological scenarios from General
Relativity coupled with perfect fluids. However, the explicit
expressions for Eqs. \eqref{eq.sol_x_func} and \eqref{eq.sol_y_func} for
obtaining specific solutions to Friedman--Lemaître--Robertson--Walker
models are very complicated.

\subsection{\ldots{} with parallel Ricci}
\label{sec:org6ded881}
A second class of solutions can be found by solving the parallel Ricci
equation, \(\nabla_\lambda R_{\mu\nu} = 0\), which yield three
independent field equations,
\begin{align}
  \nabla_t R_{tt} & \simeq \ddot{h} - ( 3 f - 2 h ) \dot{h} - h \dot{f} + 2 f h
  ( f - h ),
  \label{eq.DR_ttt}
  \\
  \nabla_i R_{ti} = \nabla_i R_{it} & \simeq  3 g \dot{h} - h \dot{g} - 2 g h (2 f -
  h) - 2 \kappa h,
  \label{eq.DR_iti}
  \\
  \nabla_t R_{ii} & \simeq \ddot{g} + g (\dot{f} + \dot{h}) + (f-h) \dot{g} - g
  h (f + h) - 4 \kappa h.
  \label{eq.DR_tii}
\end{align}
However, the system of equations is complicated enough to avoid an
analytic solution.

Despite the complication, we can try a couple of assumptions that
simplify the system of equations, for example, if one consider the
parametrisation inspired in the Friedmann--Robertson--Walker results,
i.e. setting \(f=0\), and can solve \(h\) from Eq. \eqref{eq.DR_ttt},
which is a total derivative in this particular case. Nonetheless,
despite the value of the first integration constant, the system of
equations imposes that both \(\kappa\) and \(g\) vanish.

\subsection{\ldots{} with harmonic curvature}
\label{sec:org177ec60}
Finally, the third class of solutions are those of
Eq. \eqref{eq.SimpleEOM}. The set of equations degenerate and yield a
single independent field equation,
\begin{equation}
  \nabla_\lambda R_{t i}{}^\lambda{}_{i} \simeq \ddot{g} + g \dot{f} + f \dot{g} - 2 g
  \dot{h} + 2 g h ( f - 2 h ) - 2 \kappa h
\end{equation}
Therefore, we need to set two out of the three unknown functions to be
able of solving for the connection.

\section{Conclusions and remarks}
\label{sec.concl}
In this article we have shortly reviewed the Polynomial Affine
Gravity, which is an alternative model for gravitational interactions
described solely by the connection, i.e., the metric does not play
role in the mediation of the interactions. Among the features of the
model one encounters that despite the numerous possible terms in the
action, see Eq. \eqref{eq.4dfull}, the absence of a metric tensor gives
a sort of rigidity to the action, in the sense that no other type of
terms can be added. Such rigidity suggests that if one attempts to
quantise the model, it could be renormalisable. Additionally, all of
the coupling constants, in the pure gravity regime, are dimensionless,
pointing to a possible conformal invariance of the (pure)
gravitational interactions.

Restricting ourselves to equi-affine, torsion-free connections, the
field equations are a generalisation of those from General Relativity,
Eq. \eqref{eq.SimpleEOM}. We solved the field equations for a isotropic
and homogeneous connection, either compatible with a metric or not.

When the affine connection is the Levi-Cività connection for a
Friedman--Robertson--Walker metric, we showed that the sole solution
for a Ricci flat spacetime was described by the connection of
Minkowski's space, see Table \ref{tab:metr_a_Ricci_flat}. In the
parallel Ricci case, we shown that---as intuitively expected---one
recovers the vacuum cosmological models, see Table
\ref{tab:metr_a_parallel_Ricci}, where the cosmological constant enters
as an integration constant, but such constant could be interpreted as
(partially) coming from the stress-energy tensor of a vacuum energy
perfect fluid, as mentioned---in the context of General
Relativity---in Ref. \cite{weinberg89_cosmol_const_probl}. Finally, the
(formal) solutions to the harmonic curvature are presented in Table
\ref{tab:metr_a_harm_curv}, but yet some work remains to be done to
extract physical phenomenology from these solutions.

In the case of the cosmological affine connection, we found that the
Ricci flat condition yield only two independent equations, which are
not enough to find the three unknown functions that parametrise the
homogeneous and isotropic connection. Nonetheless, since \(f\) is not
a dynamical function, the remaining two, \(h\) and \(g\), can be
solved in terms of the third one, \(f\). Interestingly, the three
functions can be chosen in a way that Ricci flatness condition for the
affine connection, yields the Friedmann-Lemaître equations from
General Relativity coupled with a perfect fluid. In this sense, the
pure Polynomial Affine Gravity supersize General Relativity, since
geometrically it can mimic effects that are usually interpreted as
\emph{matter} effects. However, among the possible solutions for the Ricci
flat condition there are countless (yet) nonphysical solutions,
and what is more, there is nothing that favours the specific choice in
Eq. \eqref{eq.fgh_param_pr} over others. Such landscape, drives us to
think that another type of condition should be use to restrict even
further the possible solutions for the affine connection.

The conditions of affine parallel Ricci could be the cornerstone in
solving the aforementioned degeneracy, since these conditions raise
three independent equations, that would serve to determine the three
unknown functions. However, at the moment we have not achieve any
interesting result in pursuing this goal.

On the other hand, the harmonic curvature condition yields a sole
(independent) field equation, and therefore the solutions are even
more degenerated than those from the Ricci flat condition, leaving
even more space for nonphysical solutions.

We would like to finish our discussion highlighting that, the geometric
emulation of matter content can serve as a starting point to a change
of paradigm related with the interpretation of the matter content of
the Universe, in particular the \emph{dark} sector.

\subsection*{Acknowledgements}
\label{sec:org0ba9562}
We would like to thank the following people for their support, without
whose help this work would never have been possible: Aureliano
Skirzewski, Cristóbal Corral, Iván Schmidt, Alfonso R. Zerwekh,
Claudio Dib, Stefano Vignolo and Jorge Zanelli. We are particularly
grateful to the developers of the software \texttt{SageMath}
\cite{stein18_sage_mathem_softw_version}, \texttt{SageManifolds}
\cite{gourgoulhon18_sagem_version,gourgoulhon15_tensor_calcul_with_open_sourc_softw,gourgoulhon18_symbol_tensor_calcul_manif}
and \texttt{Cadabra2}
\cite{peeters07_symbol_field_theor_with_cadab,peeters07_introd_cadab,peeters07_cadab},
which were used extensively along the development of this work. The
``Centro Científico y Tecnológico de Valpara'iso'' (CCTVal) is funded
by the Chilean Government through the Centers of Excellence Base
Financing Program of CONICYT, under project No. FS0821.

\appendix

\section{Lie derivative and Killing vectors}
\label{sec.Lie_der}
The usual procedure for solving the Einstein's equation is to propose
an ansatz for the metric. That ansatz must be compatible with the
symmetries we would like to respect in the problem. A first
application is seen in the Schwarzschild's metric
\cite{schwarzschild16_on_the_gravit_fiel}, which is the \emph{most general}
symmetric rank-two tensor compatible with the rotation group in three
dimensions, an thus is \emph{spherically symmetric}.

The formal study of the symmetries of the fields is accomplish via the
Lie derivative (for reviews, see
Refs. \cite{yano57_lie,choquet-bruhat89_analy,nakahara05_geomet_topol_physic,mcinerney13_first_steps_differ_geomet}). Below,
we briefly explain the use of the Lie derivative for obtaining
ansatzes for either the metric or the connection.

The Lie derivative of a connection possesses an inhomogeneous part, in
comparison with the one of a rank three tensor. This can be written
schematically as
\begin{equation*}
\mathcal{L}_{\xi} \Gamma^{a}{}_{bc} = \mathcal{L}_{\xi} T^{a}{}_{bc} + \frac{\partial^2 \xi^{a}}{\partial x^{b} \partial x^c },
\end{equation*}
or explicitly,
\begin{equation}
\label{eq.Lie_Gamma}
\mathcal{L}_{\xi} \Gamma^{a}{}_{bc} = \xi^m \partial_m \Gamma^{a}{}_{bc} - \Gamma^{m}{}_{bc} \partial_m \xi^a
+ \Gamma^{a}{}_{m c} \partial_b \xi^m + \Gamma^{a}{}_{b m} \partial_c \xi^m + \frac{\partial^2 \xi^{a}}{\partial x^{b} \partial x^c },
\end{equation}
where \(\xi\) is the vector defining the symmetry flow.

In particular, for cosmological applications, one asks for \emph{isotropy}
and \emph{homogeneity}, which in four dimensions restricts the isometry
group to either \(SO(4)\), \(SO(3,1)\) or \(ISO(3)\). The algebra of
these groups can be obtained from the algebra \(\mathfrak{so}(4)\) through
a \(3+1\) decomposition, i.e. \(J_{AB} = \set{ J_{ab}, J_{a*} }\), where the extra dimension has been denoted by an asterisk. In term of
these new generators, the algebra reads
\begin{equation}
  \label{alg-SO-decomp}
  \begin{split}
    \comm{ J_{ab} }{ J_{cd} } &= \delta_{bc} J_{ad} - \delta_{ac} J_{bd} + \delta_{ad} J_{bc} - \delta_{bd} J_{ac}, \\
    \comm{ J_{ab} }{ J_{c*} } &= \delta_{bc} J_{a*} - \delta_{ac} J_{b*}, \\
    \comm{ J_{a*} }{ J_{c*} } &= - \kappa J_{ac},
  \end{split}
\end{equation} 
with\footnote{The inhomogeneous algebra of \(ISO(n)\) can be obtained from
those of \(SO(n+1)\) or \(SO(n,1)\) through the Inönü--Wigner
contraction \cite{gilmore05_lie_group_lie_algeb_applic}.}
\begin{equation}
  \kappa = 
  \begin{cases}
    1  & SO(4) \\
    0  & ISO(3) \\
    -1 & SO(3,1)
  \end{cases}.
\end{equation}

The six Killing vectors of these algebras, expressed in spherical
coordinates are,
\begin{equation}
  \begin{aligned}
  J_1 = J_{23} &= 
  \begin{pmatrix}
  0 & 0 & -\cos\left({\varphi}\right) & \cot\left({\theta}\right)\sin\left({\varphi}\right)
  \end{pmatrix},
  \\
  J_2 = J_{31} &= 
  \begin{pmatrix}
  0 & 0 & \sin\left({\varphi}\right) & \cos\left({\varphi}\right)\cot\left({\theta}\right)
  \end{pmatrix},
  \\
  J_3 = J_{12} &= 
  \begin{pmatrix}
  0 & 0 & 0 & 1
  \end{pmatrix},
  \\
  P_1 = J_{1*} &= \sqrt{1-{\kappa} r^{2}}
  \begin{pmatrix}
  0 &  \cos\left({\varphi}\right)
  \sin\left({\theta}\right) & 
  \frac{\cos\left({\varphi}\right) \cos\left({\theta}\right)}{r} &
  - \frac{\sin\left({\varphi}\right)}{r
  \sin\left({\theta}\right)}
  \end{pmatrix},
  \\
  P_2 = J_{2*} &= \sqrt{1-{\kappa} r^{2}}
  \begin{pmatrix}
  0 &  \sin\left({\varphi}\right)
  \sin\left({\theta}\right) & \frac{
  \cos\left({\theta}\right) \sin\left({\varphi}\right)}{r} &
  \frac{ \cos\left({\varphi}\right)}{r
  \sin\left({\theta}\right)}
  \end{pmatrix},
  \\
  P_3 = J_{3*} &= \sqrt{1-{\kappa} r^{2}}
  \begin{pmatrix}
  0 &  \cos\left({\theta}\right) &
  -\frac{ \sin\left({\theta}\right)}{r} &
  0
  \end{pmatrix}.
  \end{aligned}
  \label{Kill-homotropic}
\end{equation}

Using the Eq.  \eqref{eq.Lie_Gamma}, for the above Killing vectors, the
most general connection compatible with the desired symmetries can be
obtained \cite{castillo-felisola18_beyond_einstein}, giving the
components structure shown in Eq.  \eqref{eq.ht_conn_torsion_free}.

\providecommand{\href}[2]{#2}\begingroup\raggedright\endgroup

\end{document}